\documentclass[a4paper,fleqn,usenatbib,useAMS]{mnras}

\usepackage{graphicx}	
\usepackage{amsmath}	
\usepackage{amssymb}	
\usepackage{multicol}        
\usepackage{bm}		
\usepackage{pdflscape}	

\usepackage{newtxtext,newtxmath}
\usepackage[T1]{fontenc}
\usepackage[flushleft]{threeparttable}
\usepackage[draft]{changes}

\newcommand{\gsim}{\;\lower.6ex\hbox{$\sim$}\kern-7.75pt\raise.65ex\hbox{$>$}\;}
\newcommand{\lsim}{\;\lower.6ex\hbox{$\sim$}\kern-7.75pt\raise.65ex\hbox{$<$}\;}

\newcommand{\MSUN}{$M_{\odot}$}

\newcommand {\kms} {km\,s$^{-1}$}

\newcommand{\DD}    {\mathcal{D}}
\newcommand{\LL}    {\mathcal{L}}
\newcommand{\Ri}    {R_i}
\newcommand{\Reff}  {R_{\rm e}}
\newcommand{\cnti}  {n_i}
\newcommand{\dcnti} {\delta n_i}
\newcommand{\no}    {n_0}
\newcommand{\xib}   {\boldsymbol \xi}

\title[SSH II. Recent merger or ongoing gas accretion in NGC~3741]{The Smallest Scale of Hierarchy Survey (SSH). II. Extended star formation and bar-like features in the dwarf galaxy NGC~3741: recent merger or ongoing gas accretion?}

\author[F. Annibali et al.]{
F. Annibali,$^{1}$\thanks{E-mail: francesca.annibali@inaf.it} 
C. Bacchini,$^{2}$\thanks{E-mail: cecilia.bacchini@inaf.it}
G. Iorio,$^{2,3,4}$
M. Bellazzini,$^{1}$
R. Pascale,$^{1,5}$
G. Beccari,$^{6}$
\newauthor
M. Cignoni$^{7}$
L. Ciotti,$^{5}$
C. Nipoti,$^{5}$
E. Sacchi,$^{1,8}$
M. Tosi,$^{1}$
F. Cusano,$^{1}$
S. Bisogni,$^{9}$
\newauthor
A. Gargiulo,$^{9}$
D. Paris$^{10}$
\\
$^{1}$INAF - Osservatorio di Astrofisica e Scienza dello Spazio, Via Piero Gobetti, 93/3, I-40129 - Bologna, Italy\\
$^{2}$INAF - Osservatorio Astronomico di Padova, vicolo dell'Osservatorio 5, IT-35122 Padova, Italy\\
$^{3}$Dipartimento di Fisica e Astronomia ``Galileo Galilei'', Università di Padova, vicolo dell'Osservatorio 3, IT-35122, Padova, Italy \\
$^{4}$ INFN-Padova, Via Marzolo 8, I–35131 Padova, Italy \\
$^{5}$Dipartimento di Fisica e Astronomia, Universit\`a di Bologna, via Piero Gobetti 93/2, I-40129 - Bologna, Italy\\
$^{6}$ ESO, Karl-Schwarzschild Strasse 2, D-80 Garching, Germany\\
$^{7}$Dipartimento di Fisica, Universit\`a di Pisa, Largo Bruno Pontecorvo 3, I-56127 Pisa, Italy\\
$^{8}$Leibniz-Institut fur Astrophysik Potsdam, An der Sternwarte 16, I-14482 - Potsdam, Germany \\
$^{9}$ INAF Istituto di Astrofisica Spaziale e Fisica Cosmica di Milano, Via Alfonso Corti 12, 20133 - Milano, Italy \\
$^{10}$ INAF - Osservatorio Astronomico di Roma, via Frascati 33, 00078 Monte Porzio Catone (RM), Italy 
}

\date{Accepted XXX. Received YYY; in original form ZZZ}

\pubyear{2020}

\begin{document}
\label{firstpage}
\pagerange{\pageref{firstpage}--\pageref{lastpage}}
\maketitle

\begin{abstract}

Using Large Binocular Telescope deep imaging data from the Smallest Scale of Hierarchy Survey (SSH) and archival Hubble Space Telescope data, we reveal the presence of two elongated stellar features contiguous to a bar-like stellar structure in the inner regions of the dwarf irregular galaxy NGC~3741. These structures are dominated by stars younger than a few hundred Myr and collectively are about twice as extended as the old stellar component. These properties are very unusual for dwarf galaxies in the nearby Universe and difficult to explain by hydro-dynamical simulations. 
From the analysis of archival 21-cm observations, we find that the young stellar ``bar'' coincides with an HI high-density region proposed by previous studies to be a purely gaseous bar; we furthermore confirm radial motions of a few km/s, compatible with an inflow/outflow, and derive a steeply-rising rotation curve and high HI surface density at the center, indicating a very concentrated mass distribution. We propose that the peculiar properties of the stellar and gaseous components of NGC~3741 may be explained by a recent merger or ongoing gas accretion from the intergalactic medium, which caused gas inflows towards the galaxy center and triggered star formation a few hundred Myr ago. This event may explain the young and extended stellar features, the bar-like structure, the very extended HI disc and the central HI spiral arms. The high central HI density and the steeply rising rotation curve suggest that NGC~3741 may be the progenitor or the descendant of a starburst dwarf.

\end{abstract}



\section{Introduction} \label{intro}

In the $\Lambda$ Cold Dark Matter ($\Lambda$CDM) cosmological scenario \citep{Peebles82}, galaxies are assembled over time through the accretion of smaller systems \citep{White78}. Observational evidence of this hierarchical formation process is the presence of numerous satellites and stellar streams around massive galaxies  in the Local Volume \citep[e.g.,][]{Belokurov06,McConnachie09,Martinez10,pisces,Ibata21,Malhan21}. 
Numerical simulations predict that the merger activity continues down to the lower mass scales of dwarf galaxies \citep{Diemand08,Wheeler15,Deason14};
nevertheless, dwarf-dwarf galaxy mergers have received little attention from the observational point of view so far, mostly because of the difficulty in detecting very low surface brightness merger signatures around these systems. Merger or interaction events can strongly impact the evolution of dwarf galaxies, affecting their morphology and kinematics, and providing a viable mechanism to trigger gas flows toward the inner galaxy regions and, possibly, the onset of a starburst
 \citep{Bekki08,Stierwalt15,Carlin16,Privon17,Kado20}. A number of individual dwarf-dwarf merger cases has been examined in the literature 
 \citep{Rich12,Martinez12,Sand15,Belokurov16,Amorisco14,Annibali16,Privon17,Makarova18,John19,Kallivayalil18,Zhang20}, but only a few systematic searches for dwarfs companions have been conducted so far \citep{Stierwalt15,Higgs16,Carlin16,Paduel18,Annibali20,Kado20}.  

The Smallest Scale of Hierarchy Survey \citep[SSH;][]{Annibali20} is an observational campaign designed to characterize the frequency and properties of interaction and merging events around a large sample of dwarf galaxies.  
SSH exploits  the high sensitivity and very large field of view ($\approx$23$^{\prime}$ $\times$23$^{\prime}$) of the Large Binocular Channel (LBC) on the Large Binocular Telescope (LBT). It provides deep  $g$ and $r$ photometry for 45 late-type dwarfs at distances between $\sim$1 and $\sim$10 Mpc  down to a surface brightness limit of $\mu_r\sim31$ mag arcsec$^{-2}$. 
 The SSH targets span a wide range in luminosity, from about twice the luminosity of the Large Magellanic Cloud (LMC) down to about 5 magnitudes fainter, and cover a wide range of density environments, from very isolated galaxies to group members. Photometry in two bands allows us to define the color-magnitude diagrams (CMDs) and to separate,  for targets closer than $\sim$ 4$-$5 Mpc, red giant branch (RGB) stars associated with the dwarf galaxy or with a potential satellite from background contaminants. This technique permits to reveal faint stellar substructures (e.g., streams and shells) or companions around the dwarfs. The LBC field of view translates into an explored physical region of $\sim$7$\times$7 kpc$^{2}$ to $\sim$70$\times$70 kpc$^{2}$, depending on the galaxy distance. 
 
In this paper, we present new results for the galaxy NGC~3741 (see Table~\ref{tab1} for a summary of the  main properties), which was observed as part of the SSH survey. 
NGC~3741 is a dwarf irregular (dIrr) galaxy located at a distance of $\sim$3.2 Mpc \citep[][i.e. 1\arcsec=15.7 pc]{Tully06,Dalcanton09} and has an absolute blue magnitude of $M_B=-13.2$ \citep{Cook2014}. 
NGC~3741 belongs to the Canes Venatici I galaxy cloud, which is located at the periphery of the M~81 group. NGC~3741 is supposed to be quite undisturbed by other galaxies \citep[][]{Kara03,Kara04}, being at a de-projected distance of $\simeq$1.65 Mpc from M~81 \citep{Kara02}. 

One of the most remarkable properties of NGC~3741 is the extent of the HI disc. 
Indeed, 21-cm observations with the Giant Meterwave Radio Telescope (GMRT) by \cite{Begum05} and with the Westerbork Synthesis Radio Telescope (WSRT) by \cite{Gentile07} revealed that the diameter of the HI disc is $\gtrsim14^{\prime}$ ($\sim$13 kpc) at HI column density $N_\mathrm{HI} \sim 10^{19}$ cm$^{-2}$ ($\sim 0.1$ M$_{\odot}$ pc$^{-2}$). The HI disc diameter is $\sim$21 times larger than the B-band half-light diameter, which is $\sim$40$\arcsec$ or $\sim$0.6 kpc \citep{devac91}. The total HI mass is $\mathrm{M_{HI}} \sim1.3\times10^8$ M$_{\odot}$ \citep{Gentile07,Begum08}, which is about 3.4 times larger than the stellar mass ($\mathrm{M}_\star \sim3.8\times10^7$, \citealt{Weisz11}). 

The decomposition of the rotation curve of NGC~3741 into the contribution of the individual mass components (i.e. stellar disc, gas disc, and dark matter halo) indicates that the galaxy gravitational potential is dominated by dark matter \citep{Begum05,Gentile07,Allaert2017}. The baryonic and dark matter mass estimates for NGC~3741 are consistent with scaling relations derived for dwarf galaxies \citep[e.g.][]{Thuan16,McGaugh2016,Lelli2016,iorio17,Posti2018,Mancera19,Romeo20a,Romeo20b}. 
\cite{Gentile07} analysed the kinematics of the HI disc and argued that the innermost part of the rotation curve is better reproduced by a cored dark matter profile (in which case baryons dominate the inner gravitational potential) rather than  by a cuspy one. Nevertheless, the high uncertainty in the inner rotation curve, due to the presence of non-circular motion implies that a cuspy dark matter halo cannot be definitively ruled out \citep[e.g.][]{Hayashi2006,Gentile07,Randria2015,Oman2019}.

\begin{figure*}
   \includegraphics[width=\textwidth]{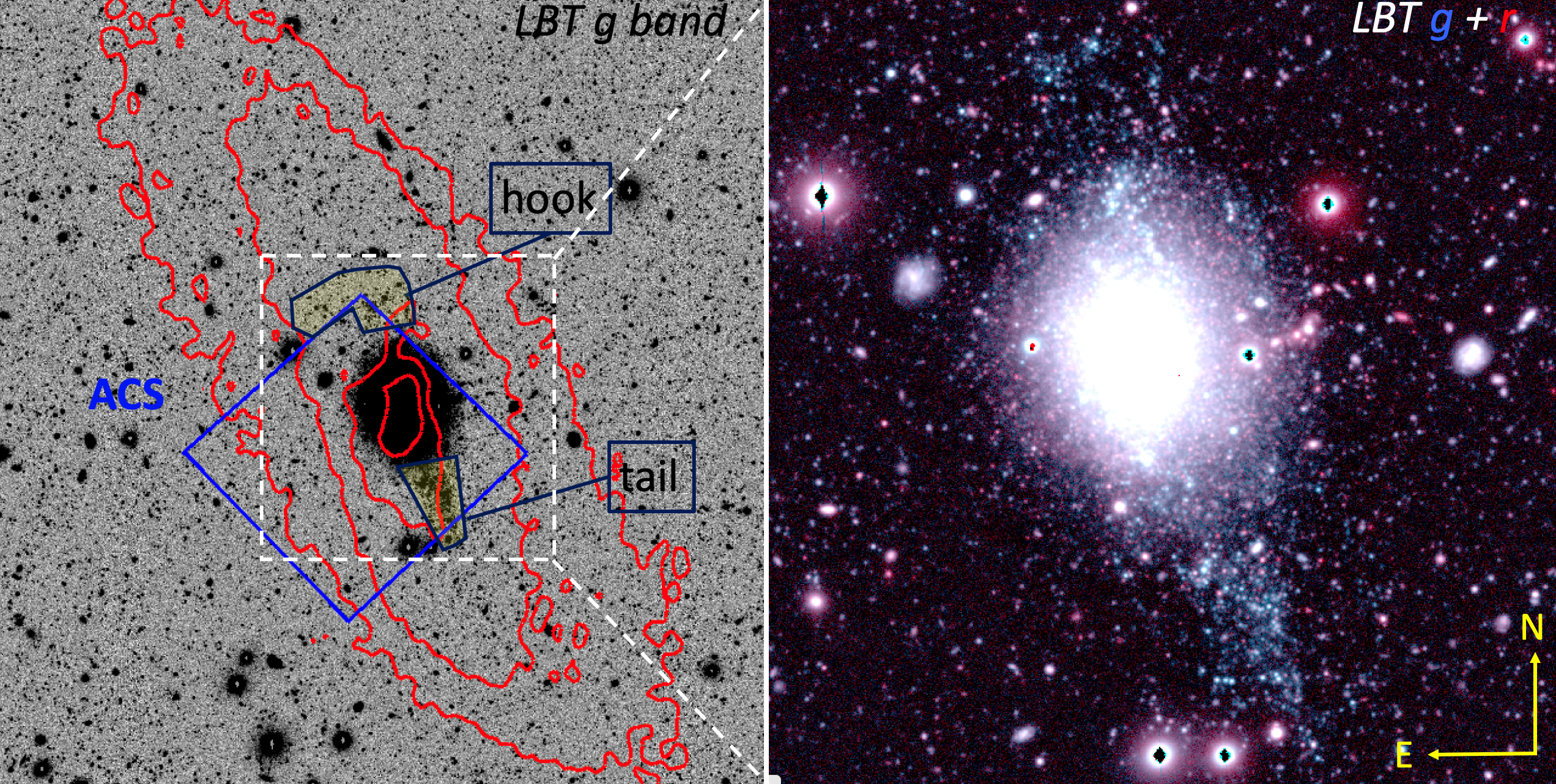} 
  \caption{Left panel: portion of the LBC $g$ image of NGC~3741 with a field of view of $\sim11^{\prime}\times11^{\prime}$, or  $\sim$10 $\times$ 10 kpc$^2$ at the galaxy distance of $\sim$3.2 Mpc. The polygons to the NE and SW of the galaxy enclose the newly identified features, dubbed the ``hook'' and the ``tail'' in this paper. The ACS footprint of archival observations includes the whole tail and a portion of the hook.  For comparison, we also show in red the contours at 0.5, 3, 8  and 15 M$_{\odot}$/pc$^{2}$ of the H~I total intensity map from the same WSRT data used in \citet{Gentile07}. Right panel: $g$ and $r$ color-combined image of NGC~3741 with a smaller field of view of $\sim4^{\prime}\times4^{\prime}$ corresponding  to the dashed square in left panel.  \label{n3741_image}
 }
\end{figure*}

Moreover, \cite{Gentile07} found that, despite the overall symmetry of the velocity field, this is distorted by non-circular motions of the order of 5-13 km/s. These motions were ascribed to the presence of a HI bar in the inner regions \citep[see also][]{Begum05,Banerjee13} and ongoing accretion of gas in the outer parts of the galaxy. Another interesting feature of the HI disc of NGC~3741 is the presence of spiral arms originating from the bar region \citep{Gentile07,Begum08,ott12}. Interestingly, while the spiral arms and the bar are visible from the atomic gas distribution, no stellar counterpart is observed \citep[e.g.,][]{Vaduvescu05}, suggesting that NGC~3741 is one of the few galaxies with purely gaseous bar and spiral arms. 
Bars may either form spontaneously as a consequence of disc instability \citep[e.g. see][for a review]{Athanassoula13b} or in response to an interaction with a galaxy companion \citep{gaida18,Pettitt18}. They can also be boosted and renewed by cold gas accretion \citep{Combes14}.  
However, the formation of purely gaseous bars remains a mystery. 
For instance, \cite{gaida18} used numerical simulations to investigate the formation of tidally-induced bars in gas-rich dwarf galaxies and found that, while bars are formed in the stellar component, no trace of the bar is found in the gaseous component.

 \begin{table}
        \centering
        \caption{Properties of NGC~3741.The last column reports the reference papers: 1=\citet{ott12}; 2=\citet{Weisz11}; 3=\citet{Lelli2016}; 4=\citet{Gentile07}; 5=\citet{Begum05}; 6=\citep{KK13}; 7=\citet{Begum08}; 8=\citet{Roy17}; 9=\citet{Berg12}.\label{tab1}}
        \label{tab:props}
        \begin{tabular}{l c c}
        \hline\hline
        Property                    & Value                                         & Ref.\\
        \hline
        Distance                    & 3.2 Mpc                                      & 1   \\
        Stellar mass                & $3.8\times10^7$ \MSUN                         & 2   \\
        Exponential disk scale length           & 0.20 kpc                                      & 3   \\
        HI mass                     & $1.3\times10^8$ \MSUN                         & 4   \\
        Rotation velocity           & 50 \kms                                       & 3, 4 \\
        Inclination                 & $68 \pm 4$°                                   & 4, 5\\
        SFR                         & $(4.3 \pm 0.7)\times10^{-3}$ \MSUN yr$^{-1}$  & 6, 7\\
        SFR surface density         & $3.2 \times 10^{-4}$ \MSUN yr$^{-1}$ kpc$^{-2}$& 8  \\
        $12 + \log(O/H)$               & $7.68 \pm 0.03$                               & 9   \\
        \hline
        \end{tabular}
\end{table}


Narrow-band H$\alpha$ and UV imaging of NGC~3741 was obtained as part of the 11 Mpc H$\alpha$ and Ultraviolet Galaxy Survey \citep[11HUGS,][]{Lee07,Kennicutt08}. 
The star formation rate (SFR) estimates derived from the H$\alpha$ and FUV luminosities are, respectively, $3.6\times10^{-3}M_{\odot}$yr$^{-1}$ and $4.9\times10^{-3}M_{\odot}$yr$^{-1}$ \citep{Begum08,KK13}, in agreement with the SFR derived from CMDs of resolved stellar populations \citep{Johnson13,Weisz11}. 
The SFR and gas surface densities averaged within the stellar disc are compatible with the Kennicutt–Schmidt and extended Schmidt laws for spirals and dwarf irregular galaxies \citep{Talbot1975,Dopita1994,Begum08,Roy15,Roy17}. 
Most of the H$\alpha$ emission is confined within a central region  of $\sim12\arcsec$ radius ($\sim$0.19 kpc), where the azimuthally averaged HI column density is $N_\mathrm{HI} \gtrsim1.7\times10^{21}$ cm$^{-2}$ ($\sim 13.6$ M$_{\odot}$ pc$^{-2}$, \citealt{Begum08}). 
However, the FUV emission extends to a larger galactocentric distance of $\approx$ 120$\arcsec$ or $\approx$ 1.9 kpc \citep[see e.g.][]{Roy17}, suggesting that star formation occurred within the last $\sim$ 100 Myr also in regions of the HI disc which are father from the galaxy center.
NGC~3741 was also observed with Spitzer in the mid- (MIR) and far-infrared (FIR) as part of the Spitzer Local Volume Legacy Survey \citep{Dale2009}. 
From FUV-to-FIR spectral energy distribution fitting, \cite{Cook2014} derived a low internal dust extintion of A$_{FUV}=0.047$ mag, in agreement with the observed trend of lower mass galaxies being less opaque than more massive ones. 
From spectroscopic observations of H~II regions, \cite{Berg12} measured an oxygen abundance of 12+ log(O/H)=7.68$\pm$0.03 (i.e. $\sim$1/12 solar metallicity), implying that NGC~3741 fits within the stellar mass-metallicity relation defined by dwarf galaxies \citep[e.g.,][]{Berg12,Pustilnik16}.

In this paper, we aim to understand the origin of the peculiar properties of NGC~3741 by comparing the stellar and the gas components of the galaxy. 
In Sect.~\ref{section_starpop}, we present the new LBT images for NGC~3741 and analyse the stellar populations using both LBT and archival Hubble Space Telescope (HST) photometry. In Sect.~\ref{sect:HI_distr_kin}, we use archival 21-cm observations to study the distribution and kinematics of the neutral gas.  Sect.~\ref{sec:discussion} provides a comparison between the properties of the stellar populations and the gas component and a discussion of the possible formation scenarios for NGC~3741. 
We give our summary in section~\ref{sec:summary}.


\section{Stellar Populations}
\label{section_starpop}

In this section, we analyse LBT and HST photometric data of NGC~3741 in order to derive the age and spatial distribution of the stellar populations using resolved-star CMDs. 

\subsection{LBT data} \label{sec:lbt_data}

   \begin{figure}
   \centering
   \includegraphics[width=\columnwidth]{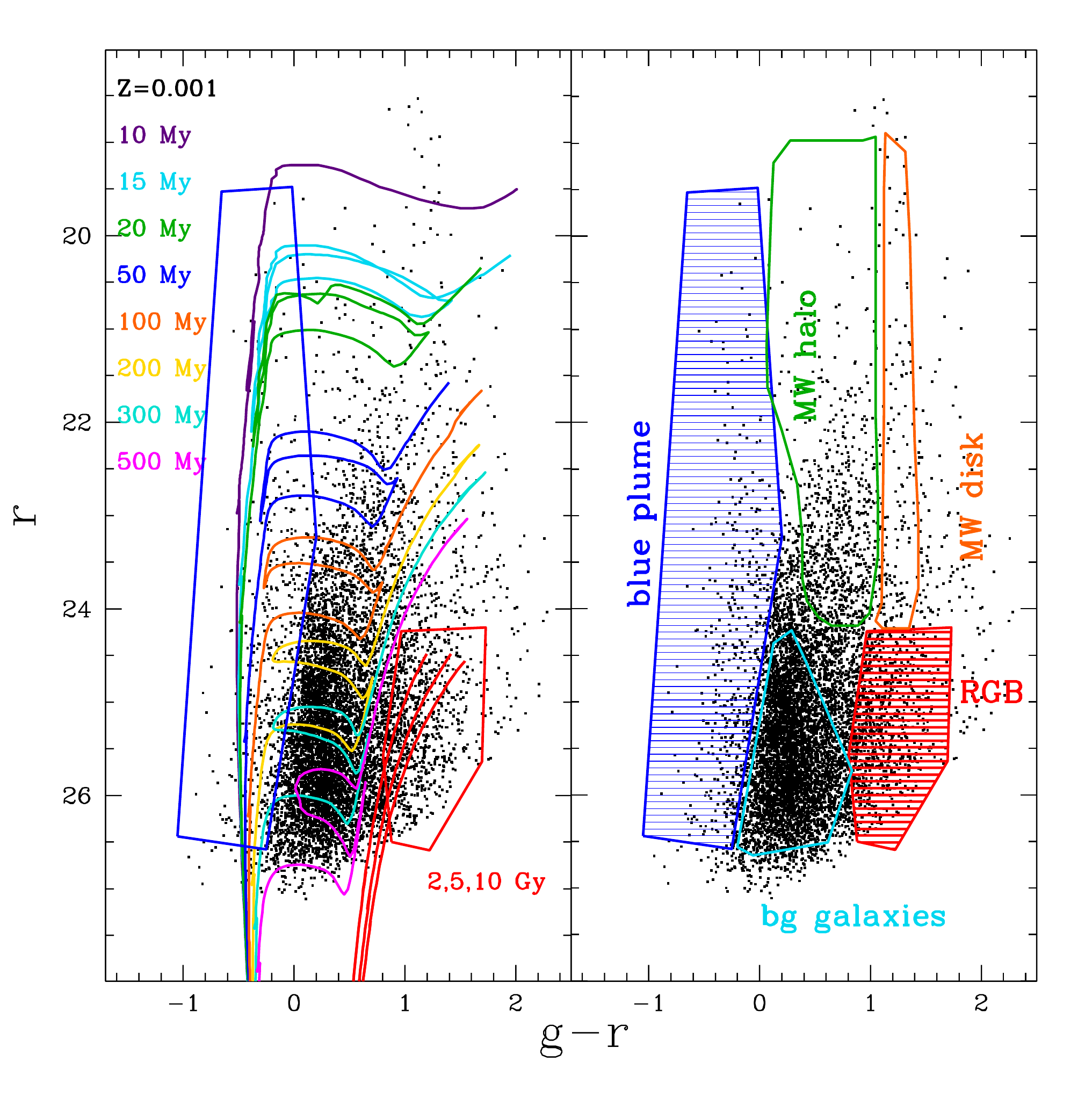}
      \caption{$r$, $g-r$ LBT CMD for sources within a $11\arcmin \times 11\arcmin$ region centered on NGC~3741. Superimposed on the left CMD are the PARSEC stellar isochrones \citep{Bressan12} in the SDSS photometric system, shifted to a distance of 3.2 Mpc and corrected for a foreground extinction of E(B-V)=0.02 from \citet{extinct}. The displayed ages are in the range 10 Myr - 10 Gyr and the metallicity is Z$=$0.001, consistent with the H~II spectroscopic metallicity of $12+\log(O/H)=7.68\pm0.03$ from \citet{Berg12}.  The polygons indicate our selection of young (age$\lesssim$300 Myr) ``blue plume'' stars and old (age$>$1-2 Gyr) RGB stars in NGC~3741, although 
     some contamination from background galaxies remains. Highlighted on the right CMD are also the regions where background galaxies, and MW halo and disc stars tend to cluster.
  \label{lbt_cmd}}
  \end{figure}

 \begin{figure*}
   \centering
   \includegraphics[width=\columnwidth]{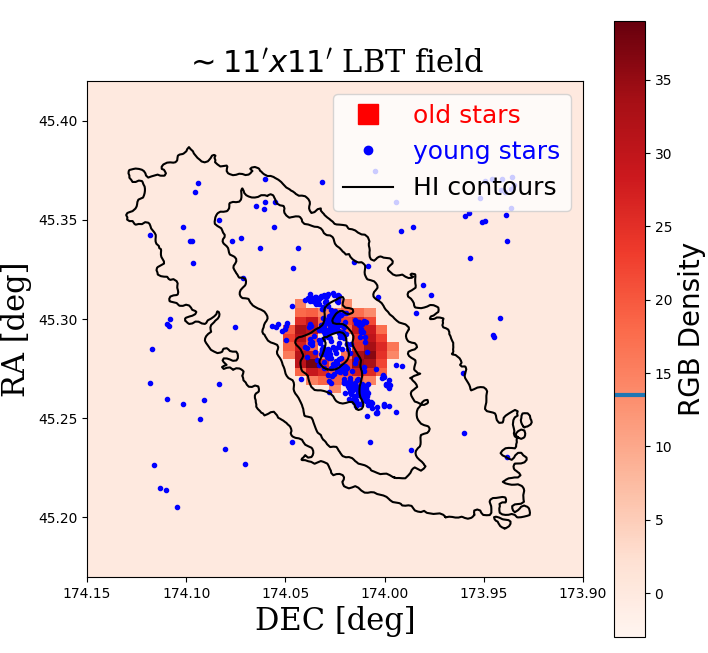}
  \includegraphics[width=\columnwidth]{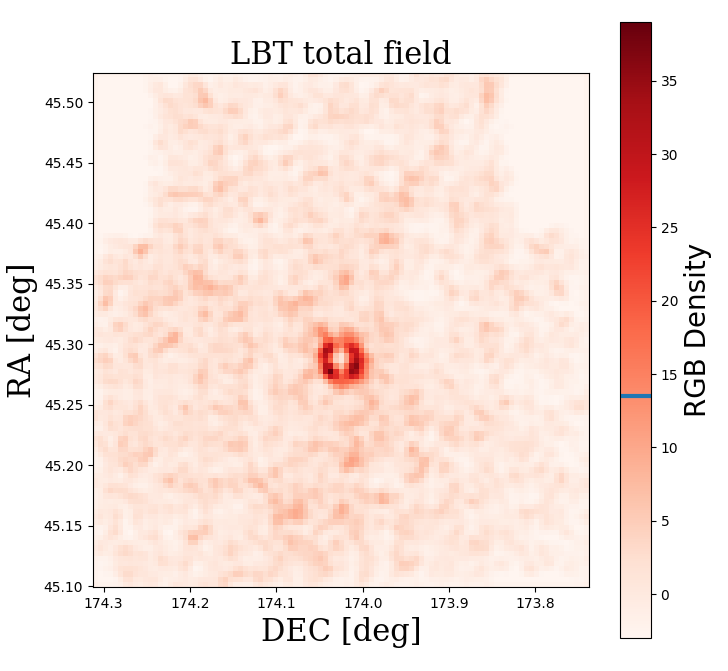}
     \caption{Spatial distribution of stars in NGC~3741 from LBT photometry. The left panel shows the density of old, RGB stars (age>1-2 Gyr) for a $11\arcmin \times 11\arcmin$ LBC region. The map is background-subtracted, and only pixels above a level of 3 times the background standard deviation (given by the horizontal blue segment in the colorbar) have been set to non-zero to mask density peaks due to red background galaxies. Blue dots are stars younger than $\lesssim$300 Myr, while the solid HI contours are the same as in  Fig.~\ref{n3741_image}.  The RGB density map for the total LBC field of view, with no pixel masking, is shown in the right panel to provide a direct visualization of the contamination from red background galaxies.  
  \label{lbt_spatial}}
  \end{figure*}

NGC~3741 was observed with the LBC on the LBT in the $g$ and $r$ bands as part of the SSH survey \citep[][hereafter Paper I]{Annibali20}.
The left panel in Fig.~\ref{n3741_image} shows a $11\arcmin\times11\arcmin$ portion of the larger $25\arcmin\times25\arcmin$ imaged field of view, 
 with superimposed the HI contours from the same WSRT data used in  \cite{Gentile07}. 
Our deep LBC data reveal the presence of two prominent blue stellar features extending in the direction of the galaxy major axis.
These features are well visible in the $4\arcmin\times4\arcmin$ $g$,$r$ color-combined image insertion displayed in the right panel of Fig.~\ref{n3741_image}: a southern triangle-shaped tail and a northern hook. 
The southern tail, which extends for  $\sim$1\arcmin  (or $\sim$0.9 kpc at NGC~3741' s distance),  
is aligned with the direction of the central HI high-density contour but is slightly offset to the west, as  we show in more detail in  Section~\ref{sect:stars_HI}.

Photometry of individual sources was performed independently on the stacked mosaic $g$ and $r$ images using PSFEX \citep{sextractor} and then the two catalogs were matched in coordinates and combined together.  Selection cuts based on the SExtractor {\it quality} flag were applied for a first removal of spurious and badly-measured objects in the photometric catalog. Then, we used diagnostics based on the comparison between aperture and point spread function fitting magnitudes to remove very extended sources, likely background galaxies, as described in details in Paper~I.

The $r$, $g-r$ CMD for sources measured within a $11\arcmin \times 11\arcmin$ region centered on NGC~3741 is shown in  Fig.~\ref{lbt_cmd}. For comparison, the PARSEC stellar isochrones \citep{Bressan12}, shifted to a distance of 3.2 Mpc and corrected for a foreground extinction of E(B-V)=0.02 \citep{extinct}, have been overplotted on the CMD. The displayed models cover a wide range in stellar ages, from 10 Myr up to 10 Gyr old. The isochrone metallicity is Z$=$0.001, which is consistent with the value expected from the oxygen abundance of the ionised gas measured by \citet{Berg12}. The CMD is heavily contaminated by background galaxies and foreground Milky Way disc and halo stars: the former mainly populate the area indicated by the cyan contours in the right panel at $-0.2\lesssim g-r\lesssim0.8$, $r\gtrsim24$, while the latter dominate the CMD region at $g-r\gtrsim0.2$, $r<24$ shown by the green and orange contours. 
As described in Paper~I, it is possible to partially separate stars belonging to NGC~3741 from foreground stars and background galaxies by selecting on the CMD the ``blue plume'' at $g-r<0$ populated by young stars (age$\lesssim$300 Myr) in the main sequence (MS) and blue core He-burning phases, and the RGB feature at $g-r\gtrsim0.8$, $r>24$ due to old (age$>$1-2 Gyr) low mass stars.

Fig.~\ref{lbt_spatial} shows the spatial distribution of stars with different ages selected from the CMD. 
In the left panel we plot the density of old, RGB stars (age>1-2 Gyr) for a $\sim11\arcmin \times 11\arcmin$ LBC region. The map was obtained by subtracting the average background computed in an external field and setting to zero all pixels below a level of 3 times the background standard deviation. This procedure allows us to mask  density fluctuations due to the unresolved background galaxies (visible in the non-masked total map in the right panel) not removed in our CMD selection.  
The result is a roundish, old stellar component; notice that the apparent lack of objects in the crowded galaxy center is an artifact of the severe incompleteness at the faintest magnitudes and must not be interpreted as a real absence of old stars in the innermost regions of NGC~3741. Bright young stars are instead detected in the galaxy center and extend beyond the old stellar component, following an elongated feature aligned in the direction of the  HI disc major axis. This is an unusual property for dIrr and blue compact dwarf (BCD) galaxies, where irregular, extended, or filamentary structures of young or intermediate-age stars are commonly present but do not encompass the old spheroid distribution \citep[see e.g.][]{Tosi01,Annibali08,Tolstoy09,Momany02,Annibali13,Higgs16,Sacchi16,Cignoni19,Annibali20}.

 \begin{figure}
   \centering
  \includegraphics[width=\columnwidth]{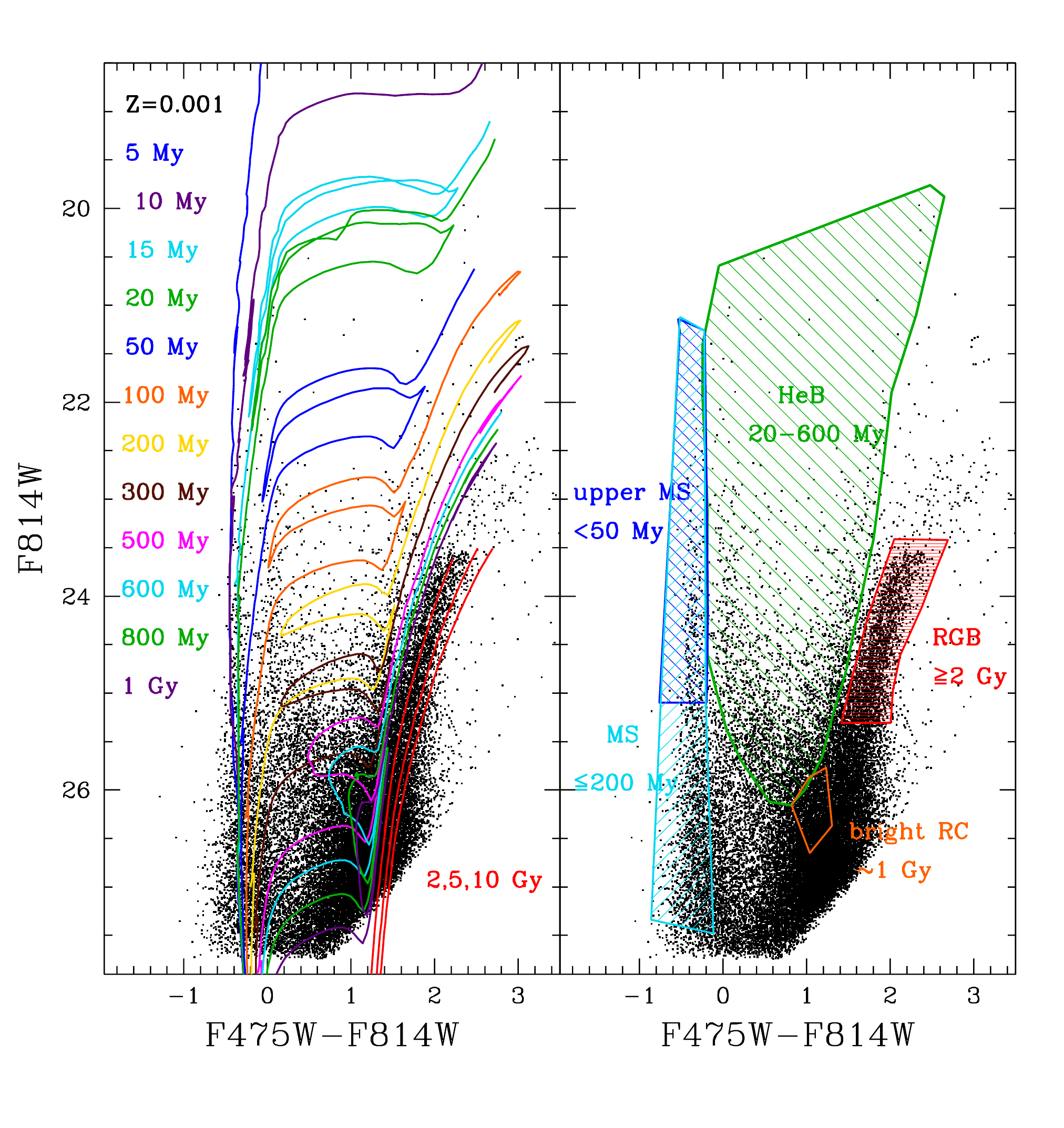}
     \caption{
F814W ($\sim$ I-band) versus F475W-F814W ($\sim$ B-I) CMD from the ACS data. In the left panel, the PARSEC stellar isochrones \citep{Bressan12} in the ACS Vegamag photometric system, shifted to a distance of 3.2 Mpc and corrected for a foreground reddening of E(B-V)=0.02, have been superimposed to the CMD. The models cover an age range from 5 Myr to 10 Gyr, as indicated in the legend, and the metallicity is Z$=$0.001. In the right panel, we highlight the five selected regions with stars in different age intervals: the MS phase with stars younger than $\sim$200 Myr; the upper MS at I$\lesssim$25 with stars younger than $\sim$50 Myr; the blue and red core He-burning (HeB) phases of massive and intermediate-mass stars with ages in the range 20-600 Myr; the brightest portion of the red clump with ages of $\sim$ 0.8-1 Gyr; a bright portion of the RGB tip with stars older than 1-2 Gyr.
  \label{acs_cmd}}
  \end{figure}

\subsection{HST data} \label{sec:hst_data}

We searched the HST public MAST archive\footnote{https://archive.stsci.edu/} for imaging data of NGC~3741. The deepest data are those acquired with the Advanced Camera for Survey (ACS) in the F475W ($\sim$B) and F814W ($\sim$I) filters  as part of the ACS Nearby Galaxy Survey Treasury \citep[ANGST,][]{Dalcanton09}. 
The calibrated photometric catalog was downloaded from the ANGRRR Photometry Repository\footnote{https://archive.stsci.edu/prepds/angrrr/datalist.html}.
The left panel in Fig.~\ref{n3741_image} shows the ACS Wide Field Camera footprint superimposed to our LBC image. The ACS pointing includes almost the entire southern tail, but it samples just a portion of the northern hook. 

The  deep ACS I, B-I CMD for  well photometred point-like sources (i.e., selected to have a Dolphot \citep{Dolphot} flag=0  and  |sharpness|$\le$0.07) is shown in Fig.~\ref{acs_cmd}.
Thanks to the high spatial resolution and depth of the ACS data, the MS and blue core-He burning phases of stars with masses $\gtrsim$ 3 M$_{\odot}$ are well separated in these CMDs. This allows for a finer age separation of the different stellar populations than with the LBT data. 
 More specifically, we identify five regions in the CMD that correspond to different age intervals: 
\begin{enumerate}
    \item the MS phase at B-I$\lesssim$-0.2 populated by stars younger than $\sim$200 Myr;
    \item a bright portion of the MS (upper MS) at I$\lesssim$25 populated by stars younger than $\sim$50 Myr; 
    \item the blue and red core He-burning (HeB) phases of massive and intermediate-mass stars with ages in the range 20-600 Myr;
    \item the brightest portion of the red clump (RC) at B-I$\sim$1.2 and 26$\lesssim$ I$\lesssim$26.5 populated by HeB stars with ages of $\sim$ 0.8-1 Gyr;
    \item a bright portion of the RGB, down to $\sim$2 mag below the tip, which allows us to isolate stars older than 1-2 Gyr.
\end{enumerate}

\begin{figure*}
   \centering
\includegraphics[width=1.\textwidth]{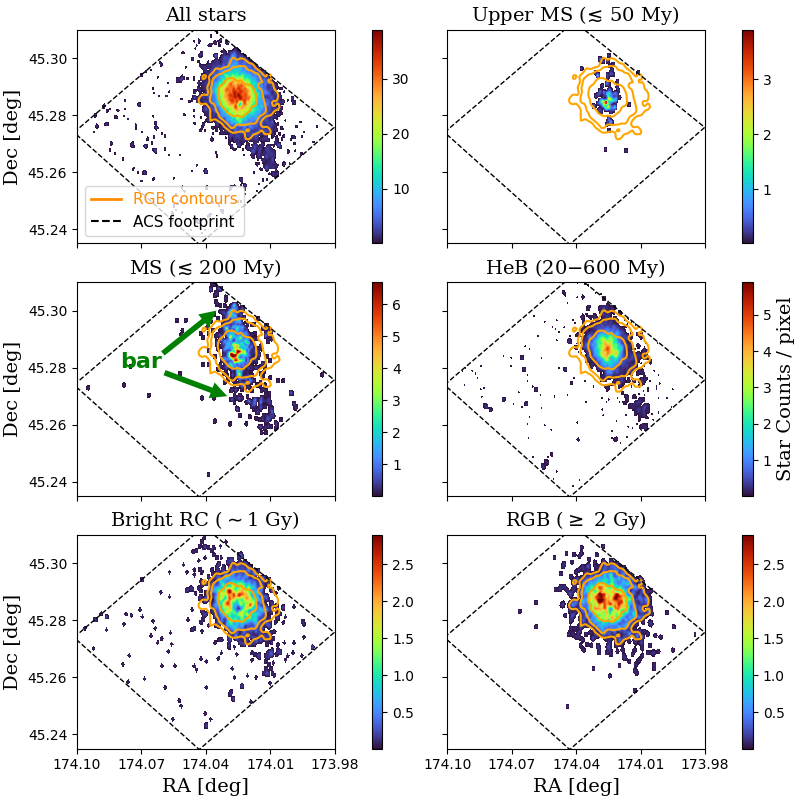}
      \caption{Stellar density maps for sources in the ACS catalog. The left upper panel shows the map 
     for the totality of measured stars, while the other panels refer to stars in different age intervals according to the CMD-based selection of Fig.~\ref{acs_cmd}: stars younger than $\sim$50 Myr, younger than $\sim$200 Myr, with ages in the 20-600 Myr range, with ages of $\sim$ 1 Gyr, and older than $\sim$2 Gyr. The contours for the old, RGB population are superimposed for reference to all spatial maps.  The dashed line denotes the ACS footprint. The identified stellar bar-like structure is indicated in the panel for ages$\lesssim$200 Myr. 
  \label{acs_spatial}}
  \end{figure*}

Figure ~\ref{acs_spatial} shows density maps for the spatial distribution of stars in these different ages bins, although some contamination from background galaxies can not be excluded also in this case.
The ACS data confirm the presence of a smooth and round-shaped distribution for the old ($\gtrsim$2 Gyr) stellar component, in agreement with the LBT data (sect.~\ref{sec:lbt_data}) and with the distribution of the 3.6$\mu$m emission (Fig.~\ref{archival}), which is a good tracer of the old stellar population.
In Appendix~\ref{app:B}, we show that the old star counts are fitted with a Sersic-profile component 
with index m$\sim$1.2, effective radius R$_e\sim$20 arcsec or $\sim$0.31 kpc \citep[see also value by][from the 3.6$\mu$m band]{Lelli2016} and M$_g\sim-12.5\pm0.5$, consistent with scaling relations derived for dwarf spheroidal galaxies \citep[e.g.,][]{Cote08,Chen10,Eigenthaler18}. 
On the other hand, very young stars ($<$50 Myr) are concentrated within a central, $\sim$0.4 kpc diameter region, as also outlined by the H$\alpha$ image from \cite{Kennicutt08} shown in Fig.~\ref{archival}.
 Stars with ages between $\sim$20 and $\sim$600 Myr, and up to perhaps $\sim$ 1 Gyr old,  present an elongated distribution and populate both the southern tail and the portion of the northern hook covered by the ACS field of view, confirming the results from the LBT data. In particular, stars younger than $\sim$200 Myr, which are very well sampled by the MS phase in the ACS CMD, exhibit a bar-like structure from which the tail and the hook appear to depart.
Also the distributions of the FUV emission and NUV emission, which respectively trace stars younger than 100 Myr and 200 Myr \citep[see][and reference therein]{KE12} resemble this "bar plus arms" configuration. Spiral galaxies often host a bar 
 and spiral arms, making NGC~3741 a sort of "young spiral galaxy", but with two important differences. The first is, of course, the stellar mass of NGC~3741, which is much lower than the typical stellar mass of spiral galaxies. The second difference is that bars in spiral galaxies are typically made of old stars.

\begin{figure*}
   \centering
   {\includegraphics[width=\textwidth]{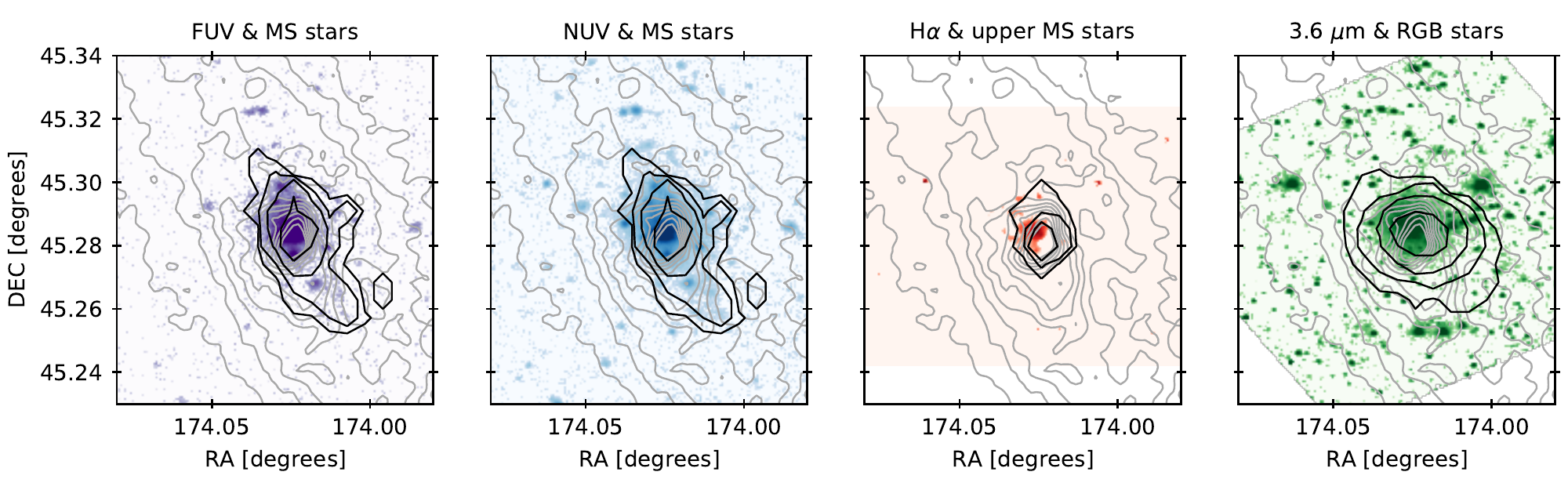}}
    \caption{Archival observations of NGC~3741: from left to right, the panels show in the FUV emission and the NUV emission observed with GALEX \citep{Lee07}, the H$\alpha$ emission \citep{Kennicutt08} and the 3.6$\mu$m emission observed with Spitzer  \citep{Dale2009}. The solid black contours indicate the distribution of different stellar populations selected from the HST CMD: MS stars with ages$\lesssim$200 Myr (first and second panels from the left), bright MS stars with ages $\lesssim$50 Myr (third panel), and RGB stars with ages > 2 Gyr (fourth panel from left). The black contours are at $2^k$ number of enclosed stars with $k=1,3,5,..19$. The light grey contours are the HI iso-density contours (see Sect.~\ref{sect:HI_distr_kin}), starting from 1~\MSUN~pc$^{-2}$ and increasing with steps of 2~\MSUN~pc$^{-2}$.} \label{archival}
  \end{figure*}

\section{HI distribution and kinematics}
\label{sect:HI_distr_kin}

We analysed the HI distribution and kinematics of NGC~3741 using the same 21-cm observations as \cite{Gentile07}, which were obtained with the WSRT (Tom Oosterloo, private communication). Because of minor differences in the data reduction, this data cube has velocity resolution of 4.1 \kms, the same as \cite{Gentile07}, but the beam size is 18.9\arcsec x 13.8\arcsec, which is slightly lower spatial resolution than their highest-resolution cube \footnote{We did also try to use the publicly available 21-cm data cubes from the survey VLA-ANGST \citep{ott12}, but we abandoned them, because we found several and relatively bright artifacts in the channels}.

In our study, we used the software $^\mathrm{3D}$Barolo\footnote{\url{https://editeodoro.github.io/Bbarolo/}} \citep{DiTeo15}, which performs a tilted-ring model fitting directly on the data cube. This software models the galaxy emission by dividing the galactic disc into a series of concentric and co-planar rings with a given width. Each ring is described by four geometric parameters (i.e. the two centre coordinates, the inclination with respect to the line of sight $i$, and the position angle PA) and four kinematic parameters (i.e. the systemic velocity of the galaxy $V_\mathrm{sys}$, the rotation velocity of the gas in circular orbits $V_\mathrm{rot}$, the gas velocity dispersion $\sigma_\mathrm{HI}$, and the radial velocity of the gas with non-circular motions $V_\mathrm{rad}$). This model is iteratively fitted to the data cube in order to find the set of free parameters that minimises the residuals between the model and the observations. Prior to the residuals minimisation, $^\mathrm{3D}$Barolo smooths the galaxy model to the same resolution of the observations, allowing to take into account the beam smearing effect. It is worth to notice that, for each ring, $^\mathrm{3D}$Barolo simultaneously fits the rotation velocity and the azimuthally averaged velocity dispersion.  This markedly improves the reliability of velocity
dispersion estimates with respect to 2D methods (e.g. 2nd-moment map of the data cube, stacking or pixel-by-pixel fitting of the line profiles) also for data with low signal-to-noise ratio (SNR).

For our modelling with $^\mathrm{3D}$Barolo, we assumed a distance of 3.2 Mpc, $V_\mathrm{sys}= 229$ \kms and the kinematic centre at (RA: 11h 36m 6.20s; DEC: 45d 17m 4.00s), which are fully compatible with the values reported in the literature \citep{Gentile07,ott12}. Prior to the model fitting, $^\mathrm{3D}$Barolo creates a mask and applies it to the data cube in order to isolate the galaxy emission. We created the mask by smoothing the data cube to a factor 2 lower resolution and selecting only the pixels with SNR$>3$ in this low-resolution cube. This procedure allows us to include also the faint galactic emission in the masked cube used for modelling the HI kinematics. 

\begin{figure*}
   \centering
   \includegraphics[width=\textwidth]{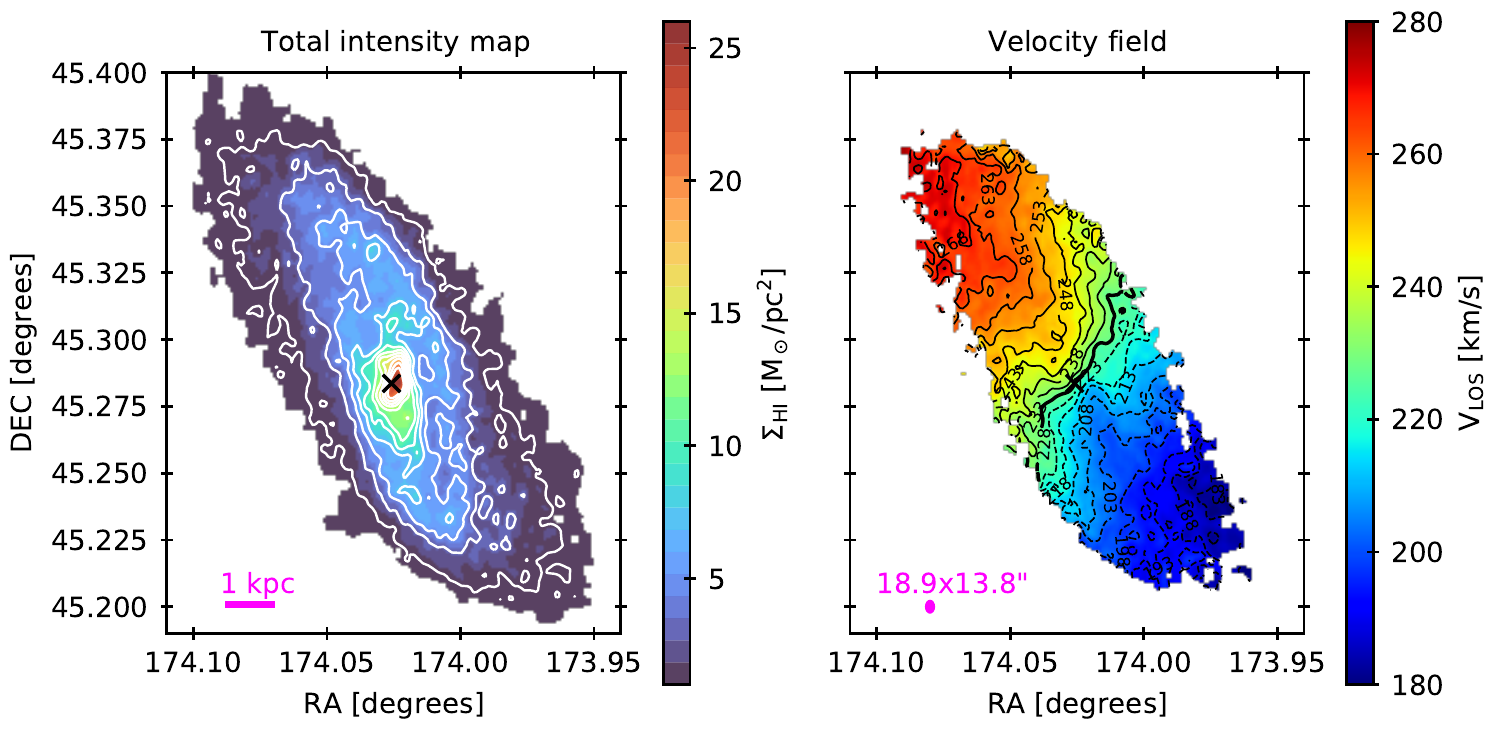}
      \caption{
      Left panel:  HI total intensity map from WSRT observations. The white contours are the iso-density curves, which start at 1~M$_\odot$pc$^{-2}$ (roughly corresponding to the $3 \sigma$ pseudo noise level in the total map; see \citealt{VerheijenSancisi2001,Lelli2014c}) and increase with steps of 2~M$_\odot$pc$^{-2}$ up to 21~M$_\odot$pc$^{-2}$. The magenta bar shows the physical scale of the observations. Right panel: velocity field (only pixels with $\Sigma_\mathrm{HI}>$1~M$_\odot$pc$^{-2}$ are shown). The black curves are the iso-velocity contours spaced by $\pm 5$~\kms from $V_\mathrm{sys}= 228$ \kms (thick contour). Solid and dashed curves are, respectively, for the receding side and the approaching side of the disc. The magenta dot shows the beam size.  In both panels, the black cross indicates the kinematic centre.
      \label{fig:2Dmaps}}
  \end{figure*}

Fig.~\ref{fig:2Dmaps} shows the total intensity map and the velocity field obtained from the masked data cube. 
From the total intensity map in the left panel, we can see that the HI disc has two evident properties: i) the HI surface density is very high in the innermost regions, where $\Sigma_\mathrm{HI} \gtrsim 26$~\MSUN~pc$^{-2}$, and  ii) inside the HI disc, two spiral arm-like structures seem to propagate from the galaxy center \citep[see also][]{Gentile07}. 
From the velocity field in the right panel, we see that the kinematic major axis (PA$\simeq 45^{\circ}$), which connects the regions with extreme line-of-sight velocities ($V_\mathrm{LOS}$), does not coincide with the geometric major axis (PA$\simeq 34^{\circ}$). Together with the distorted iso-velocity contours, this suggests the presence of a warp. 
 We note that the offset between the systemic velocity and the line-of-sight velocity along the minor axis is also a signature of radial motions \citep[e.g.][]{Fraternali2002,Lelli2012b,DiTeo21}, which might be ascribed to the presence of a bar, ongoing gas accretion, or an oval distortion of the gravitational potential \citep[e.g.][]{Bosma1978,Bosma1981,KormendyKennicutt2004}. 

To run $^\mathrm{3D}$Barolo, we fixed the width and the scale height of the rings at 15\arcsec and  10\arcsec, respectively \footnote{ $^\mathrm{3D}$Barolo is insensitive to the scale height as the tilted-ring fitting procedure is done ring-by-ring. For a thick disc, one line of sight can intersect emission from different annuli because of the projection effects due to the disc inclination. By studying the HI kinematics of dwarf galaxies, \citet{iorio17} found that assuming a constant scale height does not significantly affect the best-fit kinematical parameters, since any difference is smaller than the associated uncertainties. A value smaller than the beam is therefore a conservative choice for the scale height.}. Following \cite{Gentile07}, the galaxy inclination was fixed at $i=70^{\circ}$  for all the rings. We allowed $^\mathrm{3D}$Barolo to correct for the asymmetric drift \citep{Oh2015,iorio17}, even though the pressure support against gravity is negligible with respect to the rotational support for NGC~3741 \citep{Gentile07}.

\begin{figure*}
   \centering
   {\includegraphics[width=\textwidth]{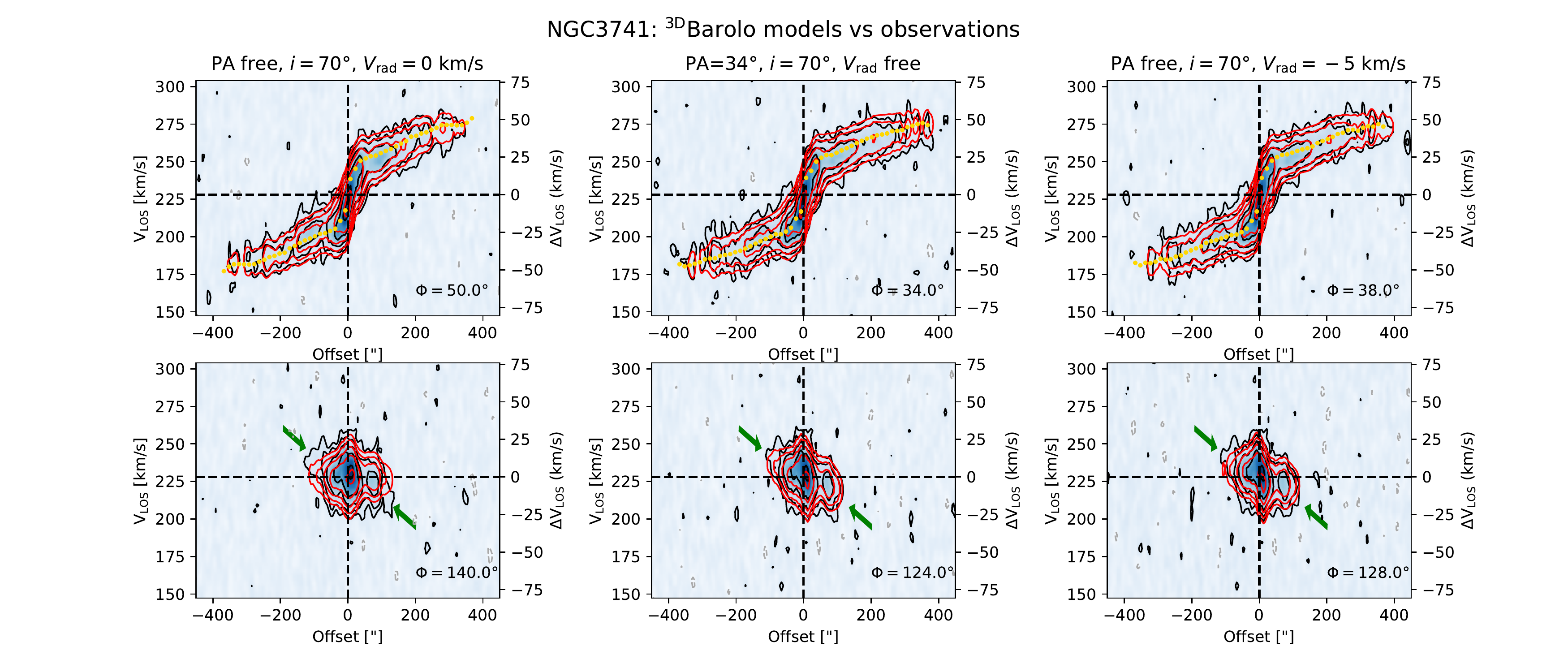}}
   {\includegraphics[width=\textwidth]{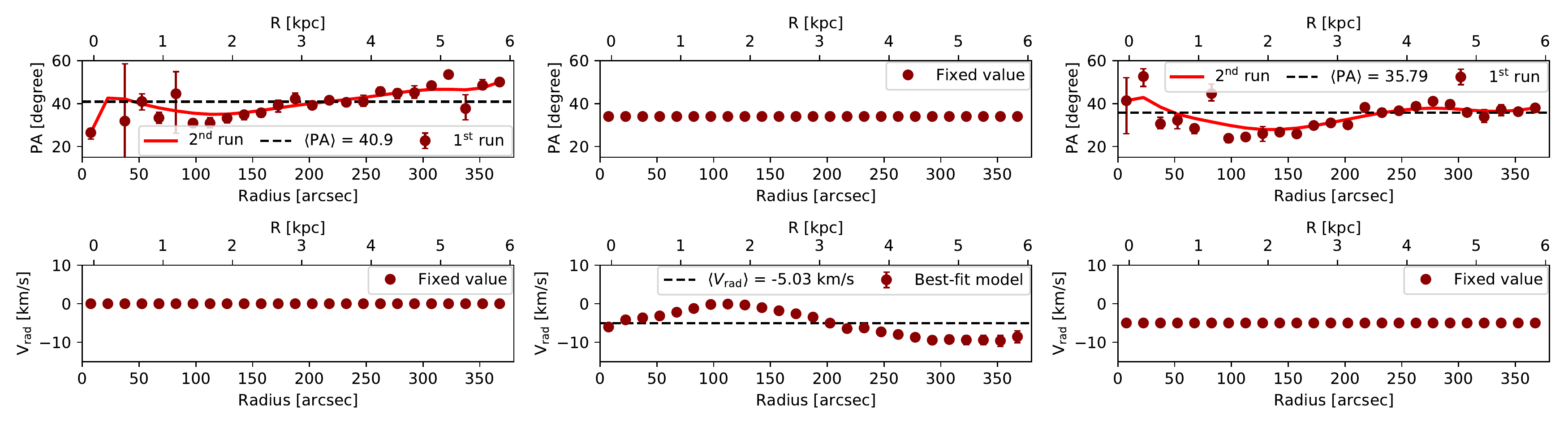}}
    \caption{Comparison between the HI observations and three best-fit models obtained with $^\mathrm{3D}$Barolo using, for each column, different sets of parameters (see text for details). \textit{1st and 2nd rows:} pv-diagram along the kinematic major and minor axes, respectively (notice that, depending on the best-fit model, the pv-diagrams are for different axis orientations, which are indicated by $\Phi$ in each panel); the observed HI emission is shown in blue (the black contours are at $2.5 \times \sigma_\mathrm{ch} \times 2^n$, with $\sigma_\mathrm{ch}=1.39$ mJy/beam being the noise in the data cube channels), while the model emission is shown by the red contours. The yellow points in the first row panels trace the rotation curve of the best-fit models. The green arrows in the second row panels indicate the emission at velocities compatible with non-circular motions. \textit{3rd and 4th rows:} PA and the radial velocity as a function of the galactocentric radius $R$ for the best-fit models. The red points show either the best-fit parameters or the assumed values, while the red curves, when present, are the regularised profiles. The black dashed line indicates the median of the free parameters. 
     \label{fig:pvmaj}}
  \end{figure*}

Since radial motions are degenerate with the viewing angles \citep{Schoenmakers1997,Schoenmakers1999}, it is not advisable to fit both the PA and $V_\mathrm{rad}$ at the same time. Hence, we first run $^\mathrm{3D}$Barolo assuming $V_\mathrm{rad}=0$ \kms, and fitting $V_\mathrm{rot}$, $\sigma_\mathrm{HI}$ and PA\footnote{In order to avoid unrealistic discontinuities, we regularised the PA radial profile by choosing the two-step fitting procedure of $^\mathrm{3D}$Barolo (\texttt{twostage=True}): after the first run, the PA profile is interpolated using a 2nd-order polynomial function, which is then used in the second run to remove a free parameter from the fit.}. The comparison between the best-fit model and the observations is provided in the left panels of Fig.~\ref{fig:pvmaj}. From the pv-diagram along the major axis (upper left panel in Fig.~\ref{fig:pvmaj}), we see that our model reproduces fairly well the HI emission. However, we notice some discrepancy in the pv-diagram along the minor axis (lower left panel in Fig.~\ref{fig:pvmaj}): although the contours of the model emission grossly reproduce the observations, there are regions (indicated by the green arrows) where the HI emission is more extended than the model contours, suggesting the presence of gas with anomalous kinematics. 
In Appendix~\ref{app:HIplots}, we show the rotation curve, the velocity dispersion radial profile and the azimuthally averaged radial profile of the HI surface density obtained for this first best-fit model.

We then built a second model by fixing the PA  at 34$^{\circ}$ based on the morphological major axis, and fitting $V_\mathrm{rot}$, $\sigma_\mathrm{HI}$, and $V_\mathrm{rad}$ (central panels, Fig.~\ref{fig:pvmaj}). Compared to the previous case, the inclusion of radial motions improves the modelling of the HI emission in the pv-diagram along the minor axis,  as indicated by the green arrows. The extent of the observed emission is well reproduced by the model. 
The radial profiles of $V_\mathrm{rot}$, $\sigma_\mathrm{HI}$ and $\Sigma_\mathrm{HI}$ of this second model are compatible within the uncertainties with the those obtained for the first model (see Appendix~\ref{app:HIplots}).

We also obtained a third best-fit model by fixing $V_\mathrm{rad}$ and fitting $V_\mathrm{rot}$, $\sigma_\mathrm{HI}$, and PA in order to include both the warp and radial motions. After various trials with different values of $V_\mathrm{rad}$, we obtained a satisfactory fit assuming $V_\mathrm{rad}=-5$ \kms (see the right panels of Fig.~\ref{fig:pvmaj}), which is consistent with the results of \cite{Gentile07}. This model reproduces fairly well the HI emission and, in particular, it is able to account for the HI with anomalous kinematics indicated by the green arrows in the pv-diagram along the minor axis. From a channel-by-channel inspection, we conclude that the second model with radial motions but no warp and the third model with both the warp and radial motions are equally good at reproducing the observations. 
 The rotation curve and the radial profiles of $\sigma_\mathrm{HI}$ and $\Sigma_\mathrm{HI}$ of this third model are compatible with those of the previous models (see Appendix~\ref{app:HIplots}).

We note that these estimates of $V_\mathrm{rad}$ can be very uncertain, since the superimposition of emission at different line-of-sight velocities can influence the determination of $V_\mathrm{rad}$. 
This issue might be important in the case of galaxies with a warp along the line of sight and for significantly thick gas discs, which is likely the case of dwarf galaxies \citep[e.g.][]{Roy10,iorio17,Patra2020,2020aBacchini}. 
Since it is not known which side of the disc is the closest to the observer, it is not possible to unambiguously discern between gas inflow or outflow, hence the sign of derived $V_\mathrm{rad}$ is also uncertain. 
If we assume that the HI spiral arms are trailing with respect to the rotation direction, we can infer that the galaxy is rotating clockwise and that the gas is inflowing with a median mass inflow rate of the order of 0.1~\MSUN/yr, in agreement with the most recent estimates for star-forming galaxies in the local Universe \citep[see][]{DiTeo21}. 

\section{Discussion} \label{sec:discussion}

In this section, we first compare the results obtained from the analysis of the stellar and the gaseous components. Then, we discuss possible evolution scenarios that can explain the peculiar properties observed in NGC~3741. 

\subsection{Comparison between stars and gas} \label{sect:stars_HI}

\begin{figure*}
\centering
\includegraphics[]{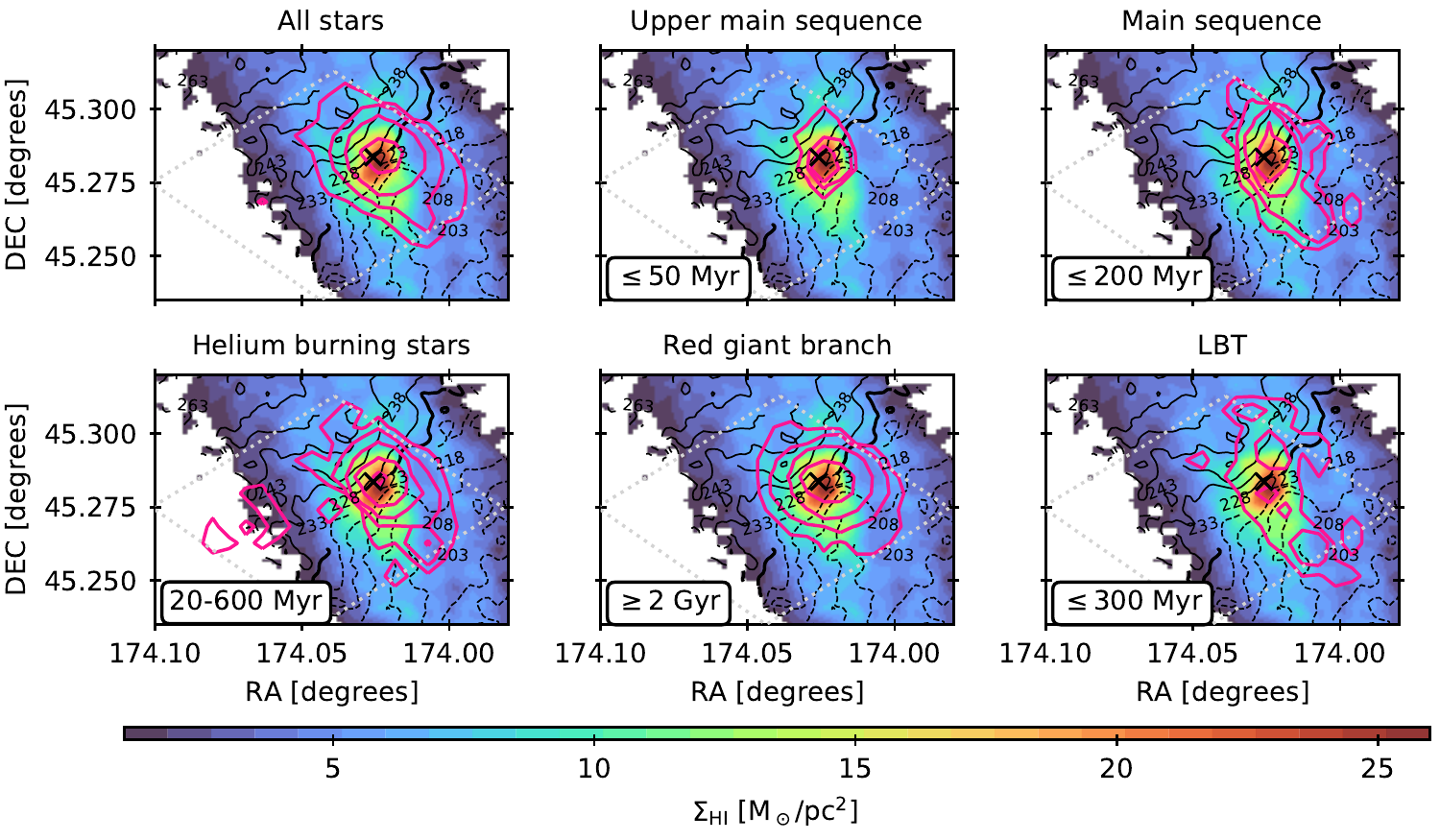}
\caption{Comparison between the distribution of stars with different ages and the HI distribution and kinematics. The HI total intensity map is colored according to the HI surface density in M$_\odot$pc$^{-2}$. The iso-velocity contours (in \kms) from the HI velocity field are shown by the solid and dashes lines for the approaching and the receding side, respectively; the thick contour indicates the systemic velocity. The grey box indicates the ACS FOV and the magenta contours show the distribution of stars (we note that the different box shape from Fig.~\ref{acs_spatial} is due to the different scale). Top left panel: distribution of all the stars from ACS data. Other panels: distribution of stars in different age bins, as indicated by boxes in the bottom left corner of the panels. The bottom right panel is obtained using the LBT observations.
\label{HI_vs_stars}}
\end{figure*}

In Figure~\ref{HI_vs_stars}, we compare the distribution of the stellar populations with different ages (see Sect.~\ref{section_starpop}) with the HI distribution and kinematics (see Sect.~\ref{sect:HI_distr_kin}), focusing on a $7\arcmin\times7\arcmin$ central galaxy region. The figure immediately emphasizes the significantly smaller spatial extension of the stellar components compared to the HI. 

The bulk of the youngest stars (age $\lesssim$50 Myr) is concentrated  within a central region of $\lesssim$1 kpc size and coincides with the highest density peak in the HI emission at $\Sigma_\mathrm{HI} \simeq 26$~M$_\odot$pc$^{-2}$.  
On the other hand, the bar-like structure, the ``northern hook'' and the ``southern tail'' ($\lesssim$300 Myr old) are aligned along the direction of the HI disc major axis and are about three times more spatially extended than the youngest stellar population. The bar coincides with the high HI density region suggested in previous studies to be a purely gaseous bar \citep[e.g.][]{Begum05,Gentile07,Banerjee13}. 
The hook and the tail appear slightly (anti-clockwise) rotated  with respect to the direction of the HI spiral arms emanating from the galaxy center. 
We notice that  the tail is located at the same position of the strong distortion in the HI iso-velocity contours at 208-213 \kms.

The oldest (age$\gtrsim$2 Gyr) stellar population is also located at the center of the HI disc, but its distribution is round and does not appear to follow the HI morphology. It encompasses regions with HI densities from  $\simeq 26$~M$_\odot$pc$^{-2}$ down to  $\simeq 6$~M$_\odot$pc$^{-2}$ with increasing galacto-centric distance. 
As discussed in Sect.~\ref{section_starpop}, this old and spheroidal stellar component appears less spatially extended than the stars with age $\lesssim$200-300 Myr. This property is unusual for dIrrs, which often host extended or filamentary structures of young or intermediate-age stars, but these are typically less spatially extended than the old stellar component \citep[see e.g.][]{Tosi01,Annibali08,Tolstoy09,Momany02,Annibali13,Higgs16,Sacchi16,Cignoni19,Annibali20}. 

\subsection{NGC~3741, a dwarf spiral galaxy?} 

The properties of NGC~3741 -- i.e. the presence of an extended, rotationally supported HI disk,  HI spiral arms, 
an old spheroidal stellar component, and extended younger stellar structures -- could suggest that we are observing a low luminosity spiral galaxy with a prominent bulge and faint spiral arms. Indeed, NGC~3741's stellar mass of $3.8\times10^7$ \MSUN is about one order of magnitude lower than the stellar mass of the smallest known spiral galaxies \citep[see e.g. the compilation in][]{Calzetti15}, which would make this system a rare, extremely low-mass spiral galaxy. 
Dwarf galaxies are known to typically lack strong spiral structures, even in the presence of an extended, gaseous disk  dominating the baryonic component.  
 \cite{Ghosh2018} suggest that in these systems the dark halo tends to suppress the growth of non axisymmetric perturbations in the gas components and that only occasional, weak spiral features (such as those observed in the HI disk of NGC 3741) can be triggered by tidal encounters or by gas accretion.

\subsection{A gaseous and stellar bar in NGC~3741?}

Previous authors \citep{Begum05,Gentile07,Begum08,Banerjee13} ascribed the presence of radial motions in NGC~3741 to a purely gaseous bar, which can be tentatively identified as a central and elongated region with high HI density (see left panel in Fig.~\ref{fig:2Dmaps}). However, purely gaseous bars are rarely observed: besides NGC~3741, the only other known cases are those of DDO~168 \citep{patra19} and NGC~2915 \citep{bureau99}. 
Indeed, the elongated, bar-like structure made of intermediate-age stars ($\lesssim$~200-300 Myr) in the top right and bottom right panels of  Fig.~\ref{HI_vs_stars} may be the stellar counterpart of the gaseous bar previously detected in NGC~3741.  
However, a bar is not identified in the stellar population older than 2~Gyr, which constitutes more than 80\% of the total stellar mass of NGC~3741 \citep{Weisz11} and exhibits a spheroidal distribution. 

Stellar bars are thought to typically form as a consequence of disc instability \citep[e.g. see][for a review]{Athanassoula13b} but several studies have shown that a dominant dark matter halo tends to slow down bars \citep{Debattista1998} or even to prevent their formation \citep{Mihos97} in dwarf galaxies.  
Stellar and gaseous bars may also form as a consequence of tidal interactions  \citep{gaida18,Pettitt18} or gas accretion \citep{Combes14}.  For instance, through $N$-body hydrodynamical simulations, \cite{gaida18} showed that stellar bars can be tidally induced by encounters with a massive host in dwarf galaxies that would otherwise be stable against bar formation for several Gyrs. The hypothesis of a tidally induced bar in NGC~3741 is appealing also because of the relatively young ages of the stars organized in the elongated,  bar-like structure, suggesting that a putative interaction may have occurred just a few hundred Myr ago. 

According to \cite{Marasco2018}, weak stellar bars can also form in dwarf galaxies if the dark matter halo is triaxial; this would also induce non-circular motions in the gas component. However, in this scenario, it seems difficult to explain the absence of a bar-like feature in the old  stellar component of NGC~3741.

\subsection{NGC~3741, the precursor/descendant of a starburst dwarf galaxy?}

 The HI analysis presented in this paper shows that NGC~3741 has a very high HI density at its centre and that the rotation curve rises steeply in the inner regions, which are properties typical of blue compact dwarf galaxies (BCDs). 
BCDs, sometimes also called star-bursting dwarf galaxies), have higher central HI surface densities than "normal" dIrrs and steeply rising rotation curves \citep[e.g.][]{Vanzee98,Simpson00,Lelli2014b}. These properties indicate a strong mass concentration at their centre, suggesting that the starburst is closely related to the central shape of the gravitational potential and to the inner concentration of gas \citep{Vanzee01,Lelli2012a,Lelli2012b,Lelli2014a,Lelli2014b}. 
Moreover, \cite{Lelli2014a} identified a population of "compact" dIrrs with steeply rising circular velocities (similar to those of BCDs) but moderate star formation activity. 
These authors proposed that compact dIrrs are the best candidates for being either the progenitors or the descendants of BCDs. NGC~3741, which shows all the typical characteristics of BCDs with the important exception of the SFR, may fit into this scenario, as discussed in the following.

 The azimuthally averaged central HI surface density of NGC~3741 is $\approx 8$ \MSUN pc$^{-2}$ (see Appendix~\ref{app:HIplots}) and its inner circular velocity gradient, defined as $V_\mathrm{c}(R_\mathrm{d})/R_\mathrm{d}$ (with $R_\mathrm{d}$ being the exponential disc scale length),  is $\sim$50 km/s kpc$^{-1}$; both these values are consistent with the typical values of BCDs and compact dIrrs \citep{Lelli2014a}. 
For comparison  DDO~87, a dIrr with HI mass and maximum circular velocity similar to those of NGC~3741, has a much lower inner HI surface density of  $\simeq 3$ M$_\odot$pc$^{-2}$ and 
a lower inner velocity gradient of $\sim$22 km/s kpc$^{-1}$.
However, the average SFR surface density within the optical radius of NGC~3741 is $\Sigma_\mathrm{SFR} \simeq 4 \times 10^{-3}$ M$_\odot$yr$^{-1}$kpc$^{-2}$ \citep{Lee07,Kennicutt08,Begum08,Johnson13,Roy17}, which is more typical of normal dIrrs rather than of strong starburst dwarfs \citep[e.g.,][]{Tolstoy09,Cignoni19}.

 What is the origin of the peculiar properties observed in NGC~3741? We suggest that a merger event or gas accretion from the intergalactic medium may be the cause. In fact, the HI disc of NGC~3741 is about seven times more extended than the stellar component, a property observed only in few other dwarfs such as DDO 154 \citep[e.g.][]{Krumm84,iorio17}, NGC~4449 \citep{Bajaja94,Lelli2014b}, NGC 2915 \citep{Meurer96} and I~Zw~18 \citep[e.g.][]{Lelli2012a,Lelli2014b}.  
It has been suggested that these extended HI discs may accumulate from the accretion of cold gas, either through minor mergers or from gaseous filaments coming from intergalactic medium  \citep[see][ and references therein]{Sancisi08}. 
In NGC~3741, this scenario is supported by the possible existence of a symmetric warp and by the presence of radial motions throughout the HI disc which seem to increase toward the galaxy outskirts.
Under the assumption that the HI spiral arms are trailing
with respect to the rotation direction, i.e. that the galaxy is rotating clockwise, the observed radial motions 
translate into an inflow of gas toward the galaxy center. 

The anomalous extended young stellar components, i.e the ``tail'', the ``hook'' and the bar-like stellar structure (superimposed to a bar-like gaseous over-density) seem to strengthen the accretion/merger hypothesis: these anomalous features may in fact originate from  gas inflow toward the galaxy center that has triggered star formation a few hundred Myr ago. However, the uncertainties in the derived SFH \citep{Weisz11} do not allow us to confirm or exclude the 
occurrence of a major starburst in NGC~3741 a few hundred Myr ago and to evaluate if its strength was comparable to those typically observed in BCDs. 
It is also possible that this galaxy is now at the beginning of a star-bursting phase and ready to turn into a BCD.

\section{Summary and conclusions} \label{sec:summary}

Our study shows that NGC~3741 exhibits peculiar properties for dwarf galaxies, both in its stellar and in its gas components. 
LBT and HST imaging revealed the presence of a bar-like stellar structure from which two elongated features, that we dub the ``northern hook'' and the ``southern tail'', appear to depart. This bar-hook-tail feature extends for $\sim$3.5 kpc in the direction of the HI disc major axis, is dominated by stars a few hundred Myr old, and is about twice as extended as the old (age>2 Gyr) stellar component. On the other hand, very young stars (age$<$50 Myr) are confined to the central ($\lesssim$1 kpc) region of the galaxy where the HI density is the highest. 
This configuration is quite uncommon among dwarf galaxies: i) although irregular, extended, or filamentary structures made of very young-to-several hundred Myr old stars are often present there, such features  do not typically encompass the spatial distribution of the old stellar component; 
ii) since their potential is dominated by the dark matter halo, dwarf galaxies are thought to be quite stable against bar formation ; iii)  
a bar composed of young stars, but not identified in the old stellar component, is quite unusual and difficult to explain.

To investigate the origin of the peculiar stellar properties and their association with the gas, we performed a new  analysis of HI archival data. The stellar bar coincides with a central, elongated region of high HI density suggested in previous studies to be a purely gaseous bar. The hook and the tail appear slightly (anti-clockwise) rotated with respect to the direction of the HI spiral arms emanating from the galaxy center. 
From the HI kinematics, we confirm the presence of HI radial motions (indicating an inflow/outflow).

The HI distribution and kinematics indicate that the surface density is very high at the galaxy centre, peaking  at $\simeq $26~M$_\odot$pc$^{-2}$, and that the rotation curve rises steeply in the inner regions, indicating a strong concentration of mass at the galaxy centre. These properties are typical of star-bursting blue compact dwarf galaxies, but less common for dwarf irregular galaxies with modest star formation rates, such as NGC~3741.

These results lead us to speculate that the unusual properties observed in NGC~3741 may be due to an advanced-stage merger with a low mass companion or to the accretion of gas from the intergalactic medium, which caused the gas to inflow towards the central regions and triggered star formation a few hundred Myr ago, forming two elongated young stellar features  (the ``tail''and the ``hook'') and a central bar-like structure superimposed to a similarly elongated HI overdensity. This accretion/interaction event may also explain the presence of a very extended HI disc and of the central HI spiral arms.  The high central HI density and the steeply rising rotation curve suggest that NGC~3741 may be the progenitor or the descendant of a starbusrt dwarf.

\section*{Acknowledgements}

We would like to thank Filippo Fraternali for the invaluable help with the data interpretation and fruitful discussions on the results, and Tom Oosterloo for providing the reduced WSRT data cube and helping with its interpretation.
We are much indebted to the anonymous referee for his/her extremely useful comments and suggestions.
We acknowledge the support from the LBT-Italian Coordination Facility for the execution of observations, data distribution, and reduction. 
MB, FA, MC and MT acknowledge the financial support from INAF Main Stream grant 1.05.01.86.28 ``SSH''. 
F. A., M.C. and M.T. acknowledge funding from INAF PRIN-SKA-2017 program 1.05.01.88.04. 

C.B. acknowledges the financial support from the European Research Council (ERC) under the European Union's Horizon 2020 research and innovation program (grant agreement No. 833824). 

Based on data acquired using the Large Binocular Telescope (LBT). The LBT is an international collaboration amongst institutions in the United States, Italy, and Germany. LBT Corporation partners are The University of Arizona on behalf of the Arizona university system; Istituto Nazionale di Astrofisica, Italy; LBT Beteiligungsgesellschaft, Germany, representing the Max-Planck Society, the Astrophysical Institute Potsdam, and Heidelberg University; The Ohio State University; and The Research Corporation, on behalf of The University of Notre Dame, University of Minnesota, and University of Virginia.

This research has made use of the SIMBAD database, operated at CDS, Strasbourg, France.
This research has made use of the NASA/IPAC Extragalactic Database (NED) which is operated by the Jet Propulsion Laboratory, California Institute of Technology, under contract with the National Aeronautics and Space Administration. 
This research has made use of NASA's Astrophysics Data System.

\section*{Data availability}

The data underlying this article will be shared on reasonable request to the corresponding author.

\appendix

\section{Projected mass profile of the old stellar component}
\label{app:B}

\begin{figure*}
\centering
\includegraphics[width=.33\textheight]{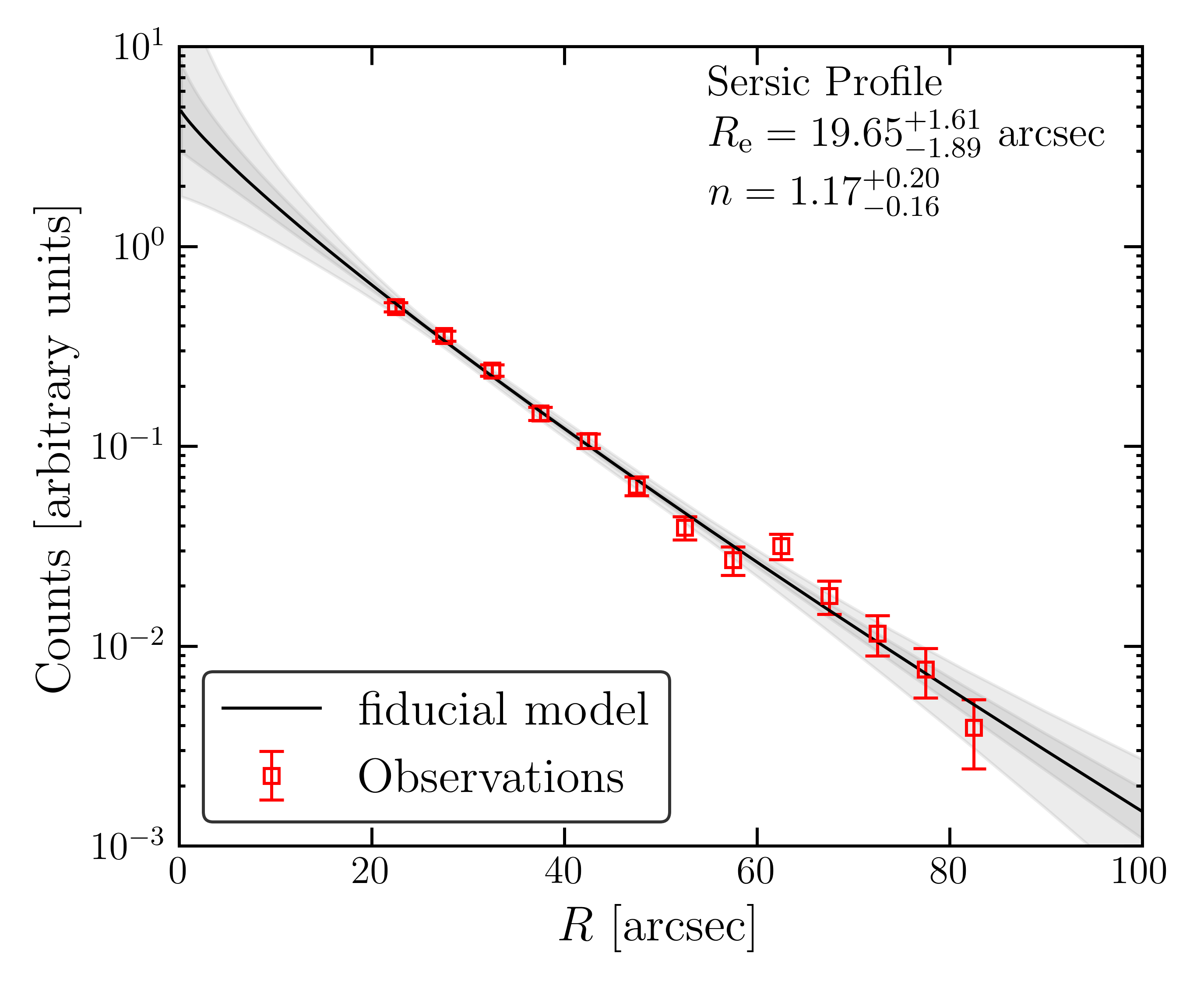}
\includegraphics[width=.33\textheight]{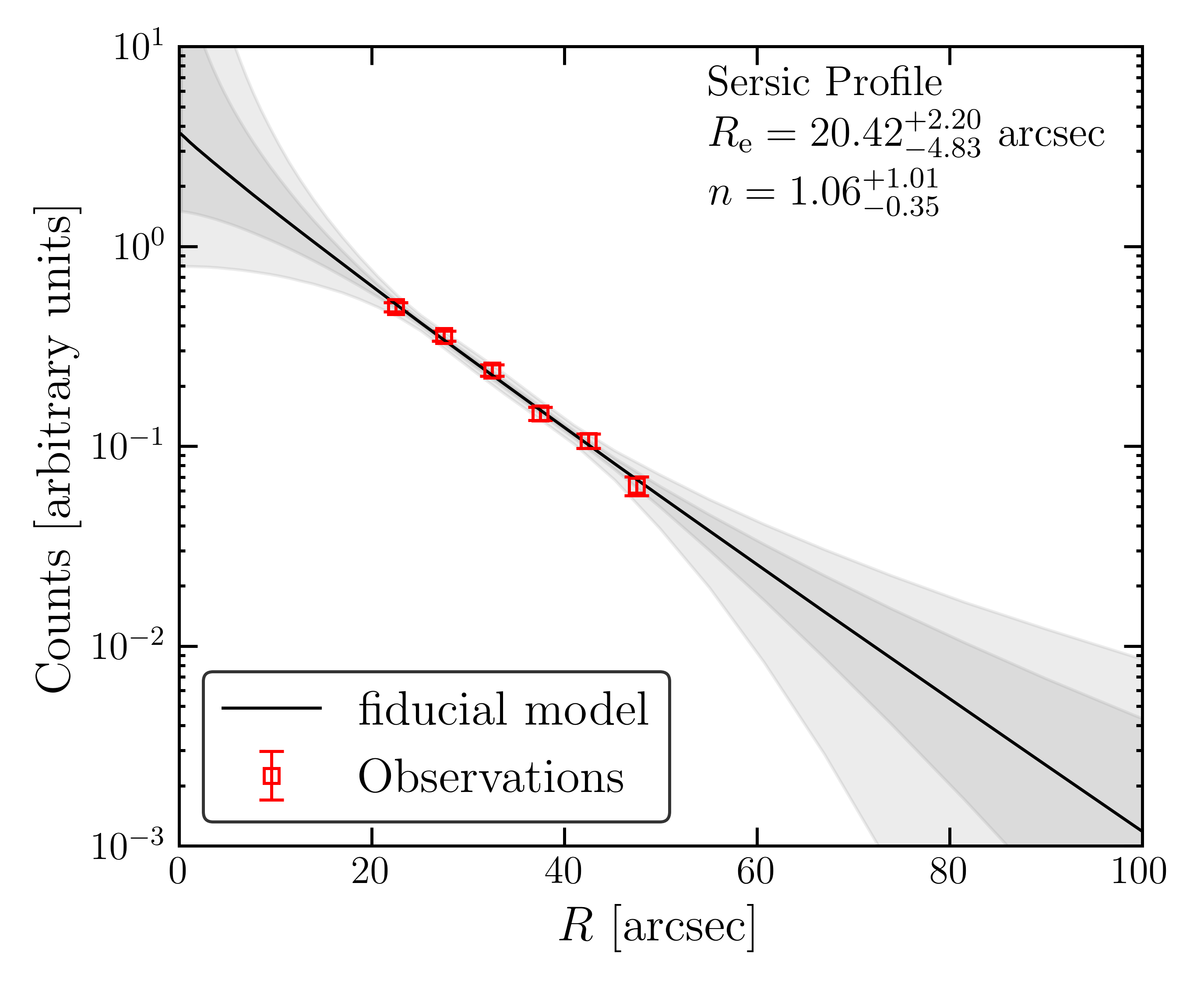}
\caption{Fit to the old component stellar profile. Red squares denote the observed RGB counts, while the solid line is our best fit. The grey shaded area delimits the 16-th and 84-th percentile region of uncertainty. The fit in the left panel was obtained considering all star counts within a galacto-centric distance of 20 arcsec < $R$ < 85 arcsec, while the fit to the right was restricted to a smaller 20 arcsec < $R$ < 50 arcsec range}. 
\label{fig:sersic_profile}
\end{figure*}

We fitted the old stellar component (projected) mass profile using individual star counts. 
To do this, we started from the HST ACS photometric catalog described in Section~\ref{sec:hst_data} and selected RGB stars down to 2 mag fainter than the RGB tip. We assume that star counts on the RGB trace the old stellar mass. We considered only stars with galacto-centric distances between 20 arcsec and 85 arcsec: star counts within 20 arcsec from the galaxy centre are highly affected by incompleteness due to severe crowding, while star counts beyond 85 arcsec can be significantly affected by background galaxies' contamination.  
The counts were binned into N=13 concentric circular annuli centred on the galaxy nominal centre. Thus the profile consists of a set of $\DD\equiv\{\Ri, \cnti, \dcnti\}$ points, with $i=1,...,N$, where $\Ri$ is the average distance of the $i$-th bin, $\cnti$ the stellar counts in that bin and $\dcnti$ the associated Poisson error. To derive the system effective radius (i.e. the distance on the plane of the sky that contains half of the stellar mass) we fit the profile with a S\'{e}rsic model
\begin{equation}
 n(R) = \no\exp\biggl[b_m \biggl(\frac{R}{\Reff}\biggr)^{\frac{1}{m}}\biggr],
\end{equation}
where $\no$ is the normalization, $\Reff$ the effective radius, $m$ the S\'{e}rsic index and $b_m$ as in equation 18 of \cite{Ciotti1999}. To explore the parameter space, we used a Markov Chain Monte Carlo (MCMC) method. The log-likelihood of the model $\ln \LL(\xib|\DD)$, defined by the parameter vector $\xib=\{\no,\Reff,m\}$, given the data $\DD$, is
\begin{equation}
 \ln \LL(\xib|\DD) = -\frac{1}{2}\sum_{i=1}^N \biggl(\frac{n(\Ri)-\cnti}{\dcnti}\biggr)^2.
\end{equation}
We ran 16 chains, each evolved for 4000 steps, we used a Metropolis-Hastings sampler \citep{Metropolis1953,Hastings1970} to sample from the posterior, and we used flat priors over the  models' free parameters. The MCMC was run by means of the \texttt{emcee} library \citep{ForemanMackey2013}. We eliminated the first 2000 steps of each chain as conservative burn-in and we used the remaining steps to build the posterior distributions over the model's free parameters. According to our fit, the estimated effective radius is $\Reff=19.65^{+1.61}_{-1.89}$ arcsec, while the S\'{e}rsic index is $m=1.17^{+0.20}_{-0.16}$, where the quoted errors have been computed as the 16-th and 84-th percentiles of the corresponding marginalized one dimensional distributions. The result of the fit is shown in the left panel of Fig.~\ref{fig:sersic_profile}. Since the more external bins may be contaminated by stars of the Tail, we tested our estimate of $\Reff$ also fitting the projected number density profile with bins in a smaller radial range of 20 arcsec < $R$ < 50 arcsec (right panel). Although with larger errors, the inference over the models free parameters is consistent with the previous case, especially for the estimate of the effective radius $(\Reff=20.42^{+2.20}_{-4.83} , m=1.06^{+1.01}_{-0.35})$.
In the end, we estimate the total magnitude of the old stellar component from the fitted profile parameters. 
To this purpose, we first perform aperture photometry on the LBT images deriving a surface brightness of $\mu_g=26.3\pm 0.3$ mag arcsec$^{-2}$ and $\mu_r= 26.5\pm0.3$ mag arccsec$^{-2}$  in an external galaxy region at $40\arcsec<R<45\arcsec$ not contaminated by young stars; then, once the Sersic profile is normalized to these values, we derive total magnitudes of $M_g=-12.5\pm0.3$ and $M_r=-12.3\pm0.3$ for the old stellar component.

\section{Radial profiles of the HI distribution and kinematics}\label{app:HIplots}

\begin{figure*}
\centering
\includegraphics[width=\textwidth]{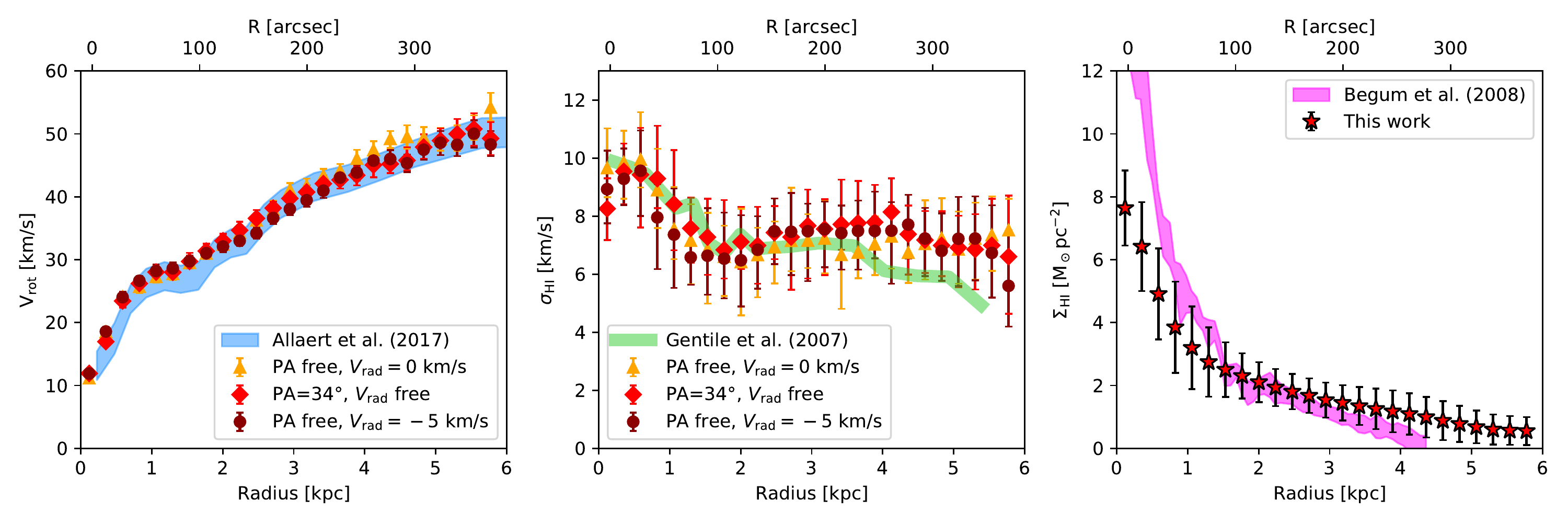}
\caption{HI rotation curve (left), radial profile of the HI velocity dispersion (centre), and azimuthally averaged radial profile of the HI surface density (right) for the three best-fit models obtained using $^\mathrm{3D}$Barolo. The  blue, green, and magenta curves in the three panels are the HI rotation curve from \citet{Allaert2017}, the HI velocity dispersion profile from \citet{Gentile07} and the HI surface density from \citet{Begum08}, respectively. 
\label{fig:vrot_vdisp_Sigma}}
\end{figure*}

This section provides the HI rotation curve, the radial profile of the HI velocity dispersion, and the azimuthally averaged radial profile of the HI surface density for the three best-fit models presented in Sec.~\ref{sect:HI_distr_kin}. 
These profiles are shown in Fig.~\ref{fig:vrot_vdisp_Sigma}, where we also provide a comparison with the HI rotation curve from \cite{Allaert2017}, the HI velocity dispersion from \cite{Gentile07}, and the HI surface density from \cite{Begum08}. 
The rotation curves and the velocity dispersion radial profiles of our models are perfectly in agreement, within the uncertainties, with each other and with the literature \citep[i.e.][]{Gentile07,Allaert2017}.
We note that, beyond $R\simeq4$~kpc, the velocity dispersion profile derived by \cite{Gentile07} from the second moment map of the WSRT data cube is slightly lower than our profiles, but the associated uncertainties are not available. 
The HI surface density obtained by \cite{Begum08} from the GMRT data cube is only grossly similar to our profile, but this could be due to the different data and masking method.

\end{document}